# Robust Dirac Point in Honeycomb Structure Nanoribbons with Zigzag Edges

*Bo Xu[1], Jiang Yin[2, *], Hongming Weng[3], Yidong Xia[1], Xiangang Wan[2, *] and Zhiguo Liu[1]*

[1]National Laboratory of Solid State Microstructures and Department of Materials Science and Engineering, Nanjing University, Nanjing, China, 210093

[2]National Laboratory of Solid State Microstructures and Department of Physics, Nanjing University, Nanjing 210093, China

[3]Research Center for Integrated Science, Japan Advanced Institute of Science and Technology, Nomi, Ishikawa 923-1292, Japan

**Abstract:**

The zigzag edge graphene nanoribbon, which is an antiferromagnetic insulator, is found from the density-function theory calculation to display a robust Dirac point after $N$ and $B$ doping at the zigzag edge. More interestingly, we found that the robust Dirac point is a common feature of the honeycomb structure nanoribbon with appropriate edge sites. The zigzag edge honeycomb nanoribbon is therefore expected to provide a very useful platform for material design and development.

[*]Authors to whom correspondence should be addressed. E-mail: jyin@nju.edu.cn, and ccmp@nju.edu.cn

The carbon-based materials have outstanding properties that make them to be interesting materials for fundamental physics and practical applications in nanotechnology.[1] Graphene, a single layer of graphite, is one of the most intriguing carbon-based materials, which has been studied intensively[2-7] since Novoselov *et al.* first succeeded in fabricating it in 2004.[2] It has been found that graphene possesses very peculiar properties such as high carrier mobility and huge coherence distance at room temperature.[3, 4] Thus it has been suggested to be a very promising candidate of the future electronic materials. Moreover, graphene has Dirac point in its band structure, and the electronic wave can propagate through the lattice and completely lose their effective mass.[3, 5] So graphene also stimulates the effort to find Dirac fermion in new materials.[8-10] It is well known that the physical properties of nanosize materials strongly depend on their geometrical structure. Therefore, to tune the physical properties by using the appropriate edge structure,[11-12] chemical substitution[13-15] and external electric field[16-18] are of both fundamental and technological importance. Cutting a monolayer graphene along a straight line will form two kinds of edges: armchair and zigzag edges. Due to the special geometric structure, graphene nanoribbons (ZGNRs) with zigzag edges have attracted particular attention.[8-33] ZGNRs have strong σ-bonds, which are far from Fermi level, and the bands around Fermi level are mainly π-like. Therefore the tight-binding Hamiltonian for the electrons can be expressed as:

$$H = \sum_i \varepsilon_i c_i^\dagger c_i - t \sum_{<i,j>} (c_i^\dagger c_j + H.c.) \qquad (1)$$

Where $c_i^\dagger$ ($c_i$) annihilates (creates) an electron at $i$ site, $\varepsilon_i$ is the on-site orbital energy at $i$ site. The sum of $<i, j>$ is taken over the nearest neighbor pairs, and $t$ is the associated hopping energy. The atoms at zigzag edge have only two nearest neighbor carbon atoms as shown in Fig 2a. Thus according to the tight-binding approximation, for these atoms located at zigzag edge, the only non-zero Hamiltonian element is:

$$H \sim (e^{i\frac{a}{2}k} + e^{-i\frac{a}{2}k}) = 2\cos\frac{a}{2}k \qquad (2)$$

where $k$ is the wave vector, and $a$ is the lattice parameter of graphene. Consequently, at the Brillouin Zone boundary, the $p_z$ orbitals of these atoms are non-bonding, thus double degenerate is just located at Fermi level. It is interesting that when the wave vector $k$ is larger than 2π/3, the band mainly contributed by the $p_z$ orbitals of two edge sites (i.e. edge state) are still located at Fermi level, thus the two-fold degenerate flat band will be formed.[34] This double degenerate partial flat band is ascribed to the non-bonding character of the $p_z$ orbitals at the edge sites. Therefore, as shown in the careful structural optimization calculation, the electron-phonon interaction can not remove the possible instability induced by the nesting behavior of the flat band. All the spin-polarized theoretical calculations predict that the flat band will induce the magnetism, and the ZGNR is an antiferromagnetic insulator with considerable magnetic moment located at edge sites.[24, 25] Usually the magnetism need correlated electrons, and it is quite rare that the magnetic moment is located at the light elements. Thus the magnetism induced by the flat-band in ZGNR attracts particular interest. Moreover, it is also found that an external transverse electric field can control this edge-induced magnetism, and results in the insulator to the half-metal transition through tuning the magnitude of the electric field. Many theoretical and experimental studies have been performed on ZGNRs for the potential applications in spintronics and nanoelectronics.[26-30]

As mentioned above, the most interesting feature of ZGNR is the two-fold degenerate partial flat band. From the rigid band point of view, if we upshift the on-site electronic energy (i.e. $\varepsilon_i$ in Eq. (1)) at the left-hand side edge sites and downshift the on-site electronic energy at the right-hand side edge sites as displayed in Fig 1a, one can expect that the two-fold degeneracy will be lifted, and the partial flat edge band by the blue line will change to the considerable dispersed band displayed by the red curve, as shown in Fig. 1b. Therefore, there exists a competition mechanism: the partial flat band may result in the magnetism, on the other hand, breaking the degeneracy of the edge sites may destroys the flat band and induces a considerable charge transfer between the edge sites. It is the bulk state band with $k$ smaller than 2π/3, and breaking the degeneracy of the edge site has only small effect on this state as indicated by the blue curve in Fig. 1b，but has significant effect on the state close to X-point. Noticing the completely different effects on the bulk state and the edge state, in addition to the transition from the magnetism to the non-magnetism, breaking the symmetry of the edge sites may also induces peculiar electronic properties at the state with the wave vector close to the border of the edge state and the body state. In this paper, we report that ZGNR becomes a nonmagnetic semiconductor and its band structure shows a robust Dirac point after *N* and *B* doping at the zigzag edges of ZGNR. More interestingly, we found that Dirac point is a quite common feature for the honeycomb structure nanoribbons with different kinds of the zigzag edges. The zigzag edge honeycomb nanoribbons are therefore expected to provide a very useful platform for material design and development.

Our electronic structure calculations are performed with the density functional theory based on the pseudo-potential plane-wave method using the implemented Vienna ab initio simulation package.[35-36] The ion-electron interactions are treated with the projected augmented wave (PAW) approximation.[37-38] The Perdew-Burke-Ernzerh functional under the spin-polarized generalized gradient correction is used to describe the exchange and correlation interaction.[39] The plane wave cutoff energy is set to 500.0 eV and the convergence threshold for energy is $10^{-5}$ eV. Brillouin Zone integration is carried out at 16×1×1 Monchorst-Pack k-grids, and 150 uniform k-points along the one-dimensional BZ were used to obtain the band structure. The symmetric unrestricted optimizations for the geometry are performed using the conjugate gradient scheme until the force acting on every atom is less than 0.01 eV/Å. The periodic boundary condition is set with the vacuum region between two neighboring nanoribbons larger than 12 Å.

As the neighbors of *C* in the periodic table, *B* and *N* are the common dopants in the carbon materials, thus we first study the effect of *B* and *N* substitution, which obviously have different electronic on-site energies with carbon. The doped ZGNR is classified by the atom number in the unit cell and the position of substitution element as shown in Fig. 2. Graphene forms the bipartite lattice where the sites can be divided into A and B sublattices. We can denote the index for the site in the *m*-th chain belonging to the A (B) sublattice as *m*A (*m*B) as shown in Fig. 2a. For example, we denote the nanoribbon with *B* atom at 1A site, *N* atom at 1B site and 18 *C* atoms at other sits as $BNC_{18}$ nanoribbon. As we know, the strength of *B-N* bond is stronger than that of *B-C,* and *N-C* bonds,[40] it is natural for us to expect that *B* and *N* atoms in the doped ZGNR will form a *B-N* pair. The existence of *B-N* pair in the doped ZGNR has been confirmed by the numerical calculations. For example, the total energy of $BNC_{18}$ nanoribbon is at least 200 meV lower than that of the nanoribbon with other atomic arrangements. Thus we designed three kinds of zigzag graphene nanoribbons doped with B-N pairs to tune the electronic on-site energy at both sides of the nanoribbons, namely: *(B-N)$_1$(C-C)$_{m-2}$(B-N)$_1$, C(B-N)$_1$(C-C)$_{m-3}$(B-N)$_1$C,* and *(CBNC)$_{m/2}$*, as displayed in Fig. 2b, 2c, and 2d, respectively. The dangling bonds at the edge of the nanoribbons as calculated are all terminated by hydrogen atoms. Fig. 2b, 2c, and 2d show the top views of the *(B-N)$_1$(C-C)$_{m-2}$(B-N)$_1$, C(B-N)$_1$(C-C)$_{m-3}$(B-N)$_1$C,* and *(CBNC)$_{m/2}$* nanoribbons with the nanoribbon width of *m*=10. The dashed rectangles indicate the unit cells, and the nanoribbons are periodic in the x-direction. Our symmetric unrestricted structural optimization shows that for three kinds of ZGNRs doped with *B-N* pairs, all atoms still form a plane. For instance, the optimized lengths of *C-N*, *B-C*, *C-C* and *B-N* bonds in the *(B-N)$_1$(C-C)$_8$(B-N)$_1$* nanoribbons are 1.39, 1.49, 1.41 and 1.44 Å, respectively. All of them show a deviation of less than 5% from the bond-length in the graphite (about 1.42 Å). Thus, we can conclude that doping *B* and *N* into the edge sites does not change the geometrical structure too much, and the structure is still much close to the perfect honeycomb lattice.

In order to find the possible magnetism, we performed several spin-polarized calculations with different magnetic arrangements as the initial state for the *(B-N)$_1$(C-C)$_{m-2}$(B-N)$_1$* nanoribbons with *m* from 2 to 30. Regardless the width of the nanoribbon and the initial spin configuration, all calculations converge to the nonmagnetic solution, and there is no magnetic moment at any sites. The nanoribbons with different widths possess the similar band structures, and the band structure of *(B-N)$_1$(C-C)$_8$(B-N)$_1$* is shown in Fig. 3a. The band close to Fermi level is still dominated by the edge sites. At X point, it is observed that the degeneracy between the lowest conduction band and the highest valence band has been lifted. The wave function of the state at the top of the valence band is located only at *N* atoms of the right edge sites and that of the state at the bottom of the conduction band is completely contributed by the *p$_z$* orbital of *B* atom of the left edge sites. In the neighborhood of *X* point, the lowest conduction band and the highest valence band show considerable energy dispersion.

The degenerate partial flat-band in ZGNR now becomes two widely dispersed bands as shown in Fig. 3a. It is the reason why *(B-N)$_1$(C-C)$_8$(B-N)$_1$* nanoribbon cannot display the magnetism. There is considerable charge transfer between B and N atoms. On the other hand, close to the zone center, the edge sites shows only small contribution, thus, as we expected, changing the edge sites does not affect these states too much, and the shape of the band close to $\Gamma$ point is similar with that of ZGNR as shown in Fig. 3a. More interestingly, we found that the edge state will cross at Fermi level at the vicinity of the border of the bulk state and the edge state, and the bands close to the crossing point can be well approximated by a linear line as shown in Fig. 3a. The band structure of *C(B-N)$_1$(C-C)$_7$(B-N)$_1$C* nanoribbon is shown in Fig. 3b. The optimized geometric structure is still close to the perfect honeycomb structure. Same as the *(B-N)$_1$(C-C)$_{m-2}$(B-N)$_1$* nanoribbons, *B* and *N* doping in *C(B-N)$_1$(C-C)$_{m-3}$(B-N)$_1$C* nanoribbons also widen the energy band at the vicinity of X point, resulting in a non-magnetic ground state. The energy-band around Fermi level can again be approximated as a linear line, and form a Dirac point. For *C(B-N)$_1$(C-C)$_{m-3}$(B-N)$_1$C* nanoribbons, all the edge sites are *C* atoms, and here we denote the edge *C* atoms close to *B* and *N* atoms as $C_B$ and $C_N$, respectively. Due to the energy difference between the $p_z$ orbitals at $C_N$ site and $C_B$ site, there is also a large charge transfer between them. The population analysis confirmed that the charge transfer is not sensitive to the width of the nanoribbon.

In addition to graphite, *BC$_2$N* is another well known layered compound [40-41] which can be synthesized by using the chemical vapor deposition with BCl$_3$ and CH$_3$CN as the starting materials.[42] Now cutting monolayer *BC$_2$N* with three elements will form rich edge configurations, which makes them interesting for material design. If we cut *BC$_2$N* sheet into zigzag nanoribbons: *(CBNC)$_{m/2}$*, as shown in Fig 2d, the ground states of *(CBNC)$_{m/2}$* nanoribbons are also non-magnetic, same as those of *(B-N)$_1$(C-C)$_{m-2}$(B-N)$_1$* and *C(B-N)$_1$(C-C)$_m$(B-N)$_1$C* nanoribbons. Fig. 4a shows a clear Dirac point-like band structure of *(CBNC)$_5$* nanoribbon. In a wide region ($3\pi/5 < \vec{k} < 4\pi/5$ or -0.5 eV < ***E*** < 0.5eV) the energy is linear-like. The charge distribution of the highest occupied molecule orbital (HOMO) and the lowest unoccupied molecule orbital (LUMO) at Dirac point in the real space is shown in Fig. 4b. It is found that Dirac point is mainly at the edge of *(CBNC)$_5$* nanoribbon, and for the other two nanoribbons, Dirac point is also mainly contributed by the edge state. All of the nanoribbons: *(B-N)$_1$(C-C)$_{m-2}$(B-N)$_1$*, *C(B-N)$_1$(C-C)$_{m-3}$(B-N)$_1$C,* and *(CBNC)$_{m/2}$* show Dirac point-like band structures with Fermi velocities of about $1.1 \times 10^5$ *m/s*, which is in one order of magnitude smaller than that of graphene. The transport of the electrons and holes in these nanoribbons should be essentially governed by Dirac's (relativistic) equation instead of the non-relativistic Schrödinger equation. Thus the charge carriers can behave like a massless Dirac fermion as in graphene.[43]

The non-magnetic ground state and Dirac point phenomenon is due to the fact that the symmetry of the edge sites is broken, thus it is very natural to expect that these properties are not restricted to the systems mentioned above. To confirm that, we performed the calculation for the zigzag-edge *BP-(C-C)$_m$-BP* nanoribbons, which have distorted honeycomb structures due to the effect of *P* atom. Our numerical results show that these nanoribbons are non-magnetic and show Dirac point near Fermi level. We also studied the zigzag edged *AlP-(Si-Si)$_m$-AlP* nanoribbons, which also have the distorted honeycomb structure. Our numerical results show that the similar band structures also exist for the zigzag-edge Si-based nanoribbons with non-symmetrical edge site.

To address the robustness of above numerical results, we performed the calculations mentioned above by using density function theory code, OPENMX,[44] which is based on the linear combination of the pseudo-atomic orbital method.[45] The high accurate double valence and the single polarization orbital were used as a basis set, which were generated by a confinement potential scheme with cutoff radii of 6.0 a.u. for B, C and N. To compare with the above results of PBE-GGA, we used the LSDA[46]. Our geometric optimization and band structure calculation confirmed that the main conclusions do not depend on the used basis set and the exchange correlation function, thus we believe that the obtained results are robust.

In summary, due to the two-degenerate partial flat band located at Fermi level, ZGNRs have drawn much attention. By using the density functional calculation, we found that B and N will form a pair when codoped into ZGNRs. Such doping almost does not change the structure, but shows the significant effect on the electronic structure. If the doping happens to break the on-site energy of the edge-sites, the system will become non-magnetism and show a Dirac point in the band structure, which will lead to peculiar electronic transport properties. Moreover, we found that the non-magnetic ground state and Dirac Ferimon behavior are the common features for the zigzag edge honeycomb nanoribbons with different on-site energies in both edge-sides. Applying the external field, such as an electric field, is one way to adjust the on-site energy. Since the electric field is not so efficient like the chemical substitution as studied in this work, the energy splitting at $X$ point and the band dispersion is not large enough to quench the nesting character. Consequently, the magnetism always appears even under an extremely large field. Another way to tune the on-site energy and break the symmetry of the edge site is to grow ZGNRs in appropriate substrates. Putting ZNGRs on BN sheet and adjusting the distance between ZGNRs and BN substrate will also lead to interesting results, which will be carried out in our future work. The states at X point will split, and Dirac point will appear.

**Acknowledgment.** The work was supported by National Key Project for Basic Research of China (Grant No. 2006CB921803, 2010CB630704, 2010CB923404 and 2010CB934201). X.G.W. also wants to thank the Fok Ying Tung Education Foundation for the financial support through Contract No. 114010.

**Figure captions:**

Fig. 1. (a) The schematic illustration of atoms at the left-hand side edge-sites with higher on-site energy $\Delta/2$; and the atoms at the right-hand side edge-sites with lower on-site energy $-\Delta/2$. The dashed rectangle indicates the unit cell. (b) The schematic picture for the band structure of the topmost valence band and the lowest conduction band. Blue line: the band dispersion for ZGNR. Red line: expected band structure for the ZGNR with the on-site energies up-shifted and downshifted at the left and right side edge-sites.

Fig. 2. (a). The structure model of graphene nanoribbon with zigzag edge. The dangling bonds at the edge are all terminated by hydrogen atoms. The top views of the $(B-N)_1(C-C)_{m-2}(B-N)_1$ (b), $C(B-N)_1(C-C)_{m-3}(B-N)_1C$ (c), and $(CBNC)_{m/2}$ (d) nanoribbons with the nanoribbon width of $m=10$. The dashed rectangles indicate the unit cells.

Fig. 3. The band structures of (a) $(B-N)_1(C-C)_8(B-N)_1$, and (b) $C(B-N)_1(C-C)_7(B-N)_1C$ nanoribbons. Fermi level is set at zero eV. The coordinates of high symmetry points are $\Gamma = 0$, and $X = \pi/a_0$. $a_0$ is the lattice parameter of the unit cell in Fig 2. The energy levels cross the Fermi level at $\vec{k} = 0.85(\pi/a_0)$, $0.7(\pi/a_0)$ for $(B-N)_1(C-C)_8(B-N)_1$ and $C(B-N)_1(C-C)_7(B-N)_1C$ nanoribbons, respectively.

Fig. 4. (a). The band structure of $(BC_2N)_5$ nanoribbon. Fermi level is set at zero eV. The energy levels cross the Fermi level at $\vec{k} = 0.65(\pi/a_0)$ for $(BC_2N)_5$ nanoribbon. (b). The charge distributions of HOMO and the LUMO at Dirac point in the real space.

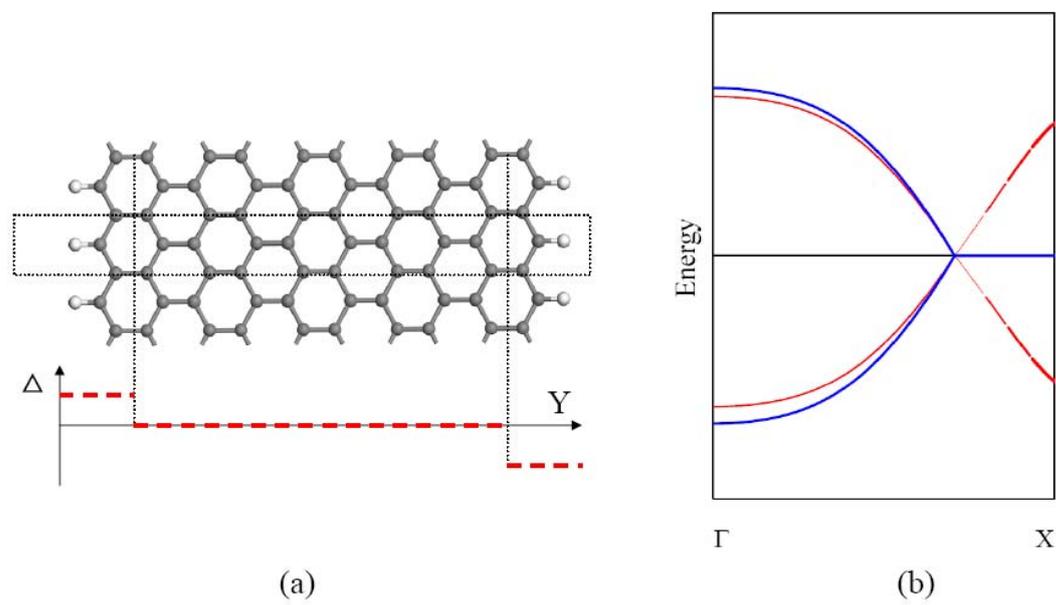

Fig. 1

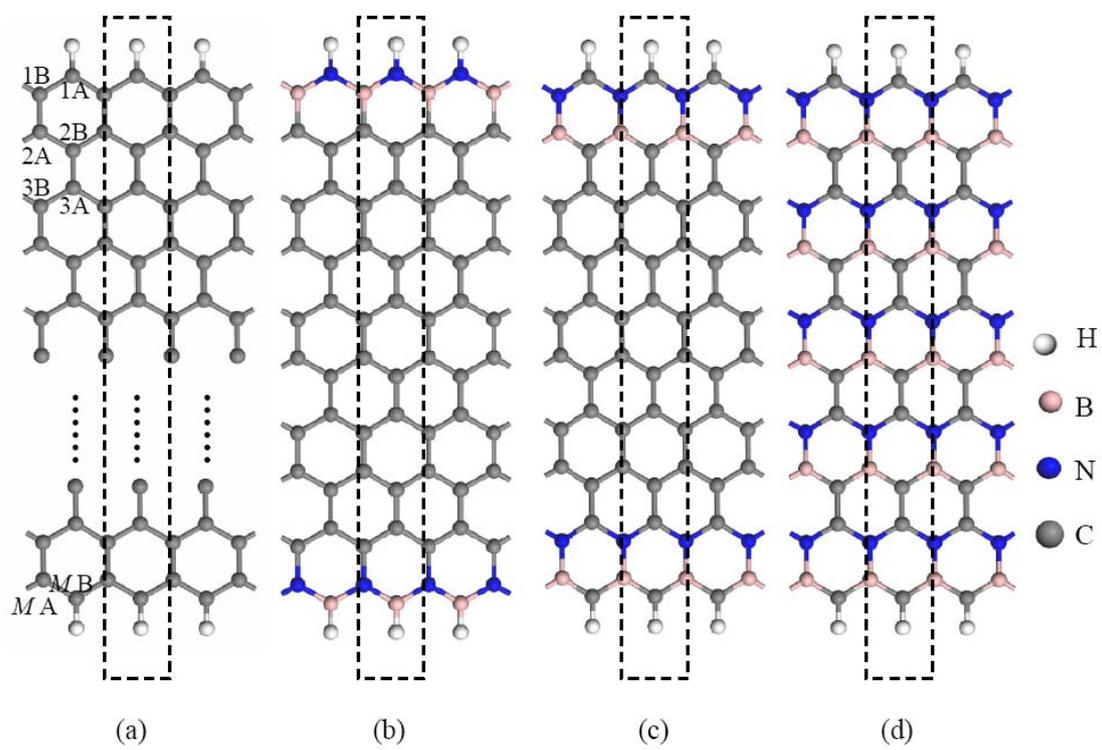

Fig. 2

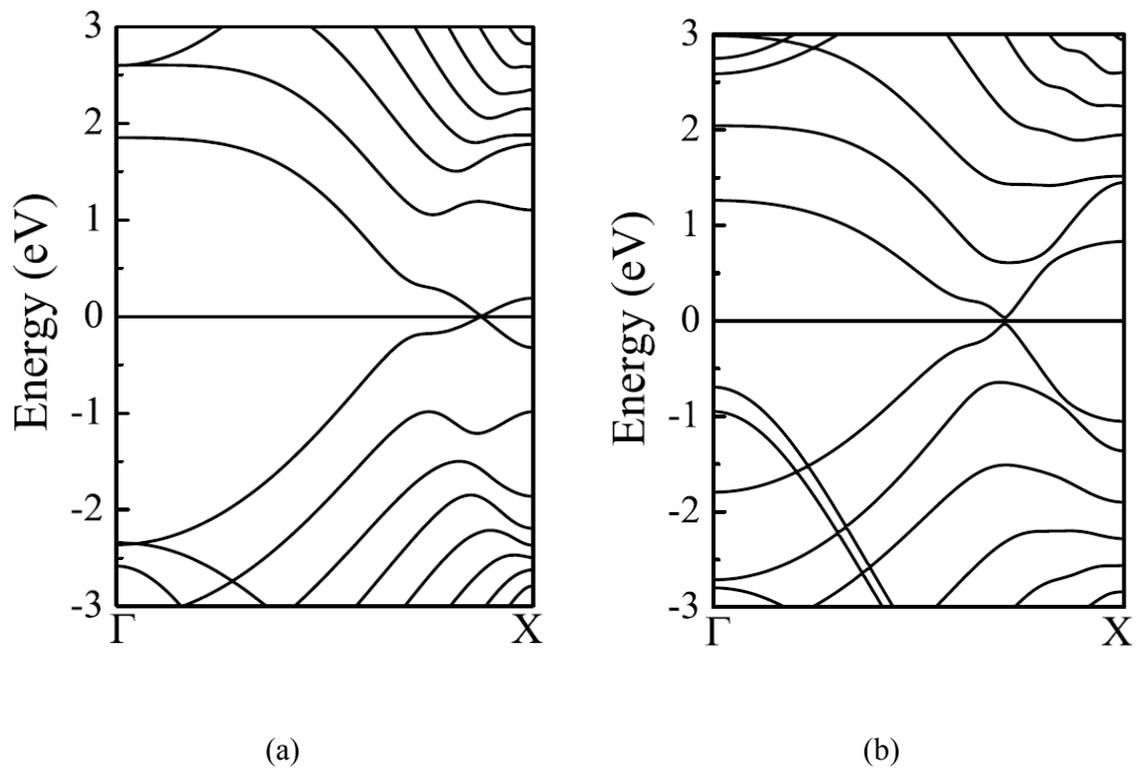

(a)            (b)

Fig. 3

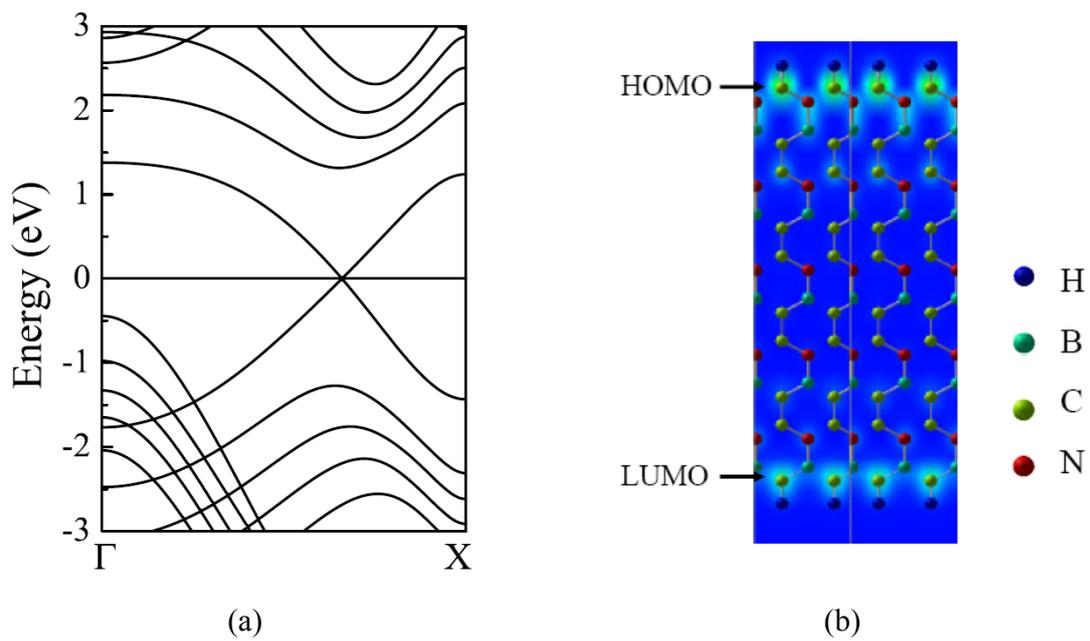

(a)　　　　　　　　　　　　(b)

Fig. 4